\documentclass[showpacs,preprintnumbers,amsmath,nofootinbib,amssymb,superscriptaddress]{revtex4}

\usepackage{caption}
\usepackage{subcaption}
\usepackage{amsmath,color}
\usepackage{amssymb}
\usepackage{bm}
\usepackage{graphicx}
\usepackage{multirow}

\usepackage{textcomp}

\allowdisplaybreaks[1]

\newcommand{\nn}{\nonumber \\}
\newcommand{\beq}{\begin{eqnarray}}
\newcommand{\eeq}{\end{eqnarray}}

\newcommand{\Slash}[1]{{\ooalign{\hfil/\hfil\crcr$#1$}}}

\usepackage{color}
\usepackage[makeroom]{cancel}
\usepackage[normalem]{ulem}
\definecolor{rbcolor}{rgb}{0.7,0.1,0}

\newcommand\rbout{\marginpar{\color{rbcolor}$\clubsuit$}\bgroup\markoverwith{\color{rbcolor}{\rule[0.4ex]{2pt}{0.8pt}}}\ULon}

\begin{document}


\title{\boldmath
$CP$-odd gluonic operators in QCD spin physics
}

\author{Yoshitaka Hatta}
\affiliation{Physics Department, Brookhaven National Laboratory, Upton, New York 11973, USA}

\affiliation{RIKEN BNL Research Center,  Brookhaven National Laboratory, Upton, New York 11973, USA}


\date{\today}

\begin{abstract}

We explore connections between high energy QCD spin physics and  $CP$-odd scalar gluonic operators  $\tilde{F}^{\mu\nu}F_{\mu\nu}$ and  $\tilde{F}_{\mu\nu}F^{\mu\alpha}F^{\nu}_{\alpha}$, the latter being  called the Weinberg operator in the context of the nucleons' electric dipole moment.  We first introduce the twist-four generalized parton distribution (GPD) associated with the topological operator $F_{\mu\nu}\tilde{F}^{\mu\nu}$. This has interesting applications in spin physics which go beyond the 
 standard framework in terms of twist-two and twist-three distributions.  
In the second part, we show that 
the off-forward matrix element of the  Weinberg operator  is proportional to a certain twist-four correction to the $g_1$ structure function in polarized deep inelastic scattering.

\end{abstract}


%
%
%

\maketitle

%
%
%

\section{Introduction}

This paper is a natural sequel to the previous work \cite{Hatta:2020iin} which discussed the parton distribution function (PDF) associated with the scalar gluonic operator 
\beq
F(x) \sim \int dz^- e^{ixP^+z^-}\langle P|F^{\mu\nu}(0)F_{\mu\nu}(z^-)|P\rangle, \label{ff}
\eeq
where $|P\rangle$ is the nucleon single particle state. 
The main motivation of \cite{Hatta:2020iin}  was to study the partonic structure of the nucleon mass. Eq.~(\ref{ff}) is suitable for this purpose because the first moment $\int dx F(x)\sim \langle P|F^{\mu\nu}F_{\mu\nu}|P\rangle$ is proportional to the `gluon condensate' which is related to the nucleon mass via the QCD trace anomaly. Being a twist-four distribution, $F(x)$ affects experimental observables only at the subleading order in the usual twist expansion. Yet it can provide fundamentally important insights into our understanding of the origin of hadron masses in QCD, a problem recently proclaimed as one of the major goals of the Electron-Ion Collider (EIC) \cite{nas}. 

A simple variant of (\ref{ff}) is another twist-four PDF, or more precisely,  generalized parton distribution (GPD) 
\beq
\tilde{F}(x,\Delta) \sim \int dz^- e^{ixP^+z^-}\langle P'|\tilde{F}^{\mu\nu}(-z^-/2)F_{\mu\nu}(z^-/2)|P\rangle,  \label{ftilde}
\eeq
 whose first moment gives the nucleon matrix element of the $CP$-odd scalar gluonic operator $F^{\mu\nu}\tilde{F}_{\mu\nu}$. It is necessary to introduce nonvanishing momentum transfer  $P'-P=\Delta\neq 0$, or else the distribution vanishes. Just as $F(x)$ is related to the partonic structure of the nucleon mass, $\tilde{F}(x)$ is related to that of the nucleon spin---another major goal of the EIC.  The original motivation of this paper was to  explore this connection which goes beyond the standard description of the nucleon spin structure in terms of twist-two and twist-three distributions. Of course, the relevance of  the  operator $F\tilde{F}$ to  QCD spin physics is by no means novel.   There is a decade-long controversy over the role of  $F\tilde{F}$ in the nucleon spin puzzle through the chiral anomaly, see, e.g., \cite{Jaffe:1989jz,Cheng:1996jr}. However, most of the discussion in the literature is concerned with the integrated (local) operator $F\tilde{F}$, with a notable exception in \cite{Mueller:1997zu}. It is interesting see whether  nonlocality in the light-cone direction can bring about new insights into the problem. 
 Indeed, very recently, Tarasov and Venugopalan \cite{Tarasov:2020cwl} have  identified precisely the same distribution (\ref{ftilde}) in their `worldline' approach to box diagrams for the $g_1$ structure function in polarized deep inelastic scattering (DIS). In view of such developments,  it is timely to study the general properties of   $\tilde{F}(x)$ from QCD perspectives. 
 
In addition to the above project  which has many parallels to the analysis done in \cite{Hatta:2020iin}, we have noticed that introducing the $x$-dependence in the $F\tilde{F}$ sector may open up a new research direction of interdisciplinary nature.  
In a recent paper \cite{Seng:2018wwp}, Seng  suggested that the third moment of the twist-three, chiral-odd PDF $e(x)$ is related to the so-called quark  chromo-magnetic dipole moment (cMDM) operator important in the context of $CP$-violation  and the electric dipole moment (EDM) of the nucleons. This is an interesting new connection between nucleon structure studies and physics beyond the Standard Model (BSM).   We point out that an entirely analogous connection exists between   the third moment of $\tilde{F}(x)$ and   the so-called Weinberg operator \cite{Weinberg:1989dx}
\beq
  f_{abc}\tilde{F}_{\mu\nu}^a F^{\mu\alpha}_b F^{\nu}_{c\alpha}, \label{op}
\eeq
which is another candidate operator to generate a large EDM in the nucleon. Based on this observation, we establish a relation between the matrix element of the Weinberg operator and observables in spin physics.  This suggests an exciting possibility that polarized DIS experiments   can provide useful information to the physics of the nucleon EDM, or more generally, BSM-origins of hadronic $CP$ violations. 

This paper is structured as follows. In Section II, we give a precise definition of (\ref{ftilde}) and discuss its connection to the chiral anomaly and the nucleon spin decomposition. In Section III, we perform one-loop calculations of $\tilde{F}(x)$ for quark and gluon targets. In Sections IV and V, we discuss the third moment of $\tilde{F}(x)$ and identify an operator similar to the Weinberg operator. Through a detailed  analysis of the properties of these operators, we shall derive a relation between the off-forward matrix element of the Weinberg operator and one of the twist-four corrections to the $g_1$ structure function in polarized DIS. 
 
 \section{Generalized parton distribution of $F\tilde{F}$}

We start by defining the twist-four gluon GPD for the nucleon 
 \beq
 \tilde{F}(x,\xi,\Delta^2) &\equiv & \frac{-i\bar{P}^+}{2M^2}\int \frac{dz^-}{2\pi} e^{ix\bar{P}^+z^-} \langle P'S'|\tilde{F}_{\mu\nu}(-z^-/2) WF^{\mu\nu}(z^-/2)|PS\rangle \nn 
 &=& \frac{1}{2M^2}\bar{u}(P'S')\left[H^+\Delta^+\gamma^- + H^-\Delta^-\gamma^+ -H_\perp\Delta^i_\perp \gamma^i_\perp \right]\gamma_5 u(PS),
 \eeq
 where $M$ is the nucleon mass and $\Delta=P'-P$ is the momentum transfer. $S^\mu$ is the spin four-vector which satisfies $S\cdot P=0$ and $S^2=-M^2$.  We work in a frame in which $\bar{P}=\frac{P+P'}{2}$ has vanishing transverse components.  $W$ is the light-like adjoint Wilson line  which makes the operator gauge invariant (and will be often omitted in the following). 
  Our convention is  $\epsilon^{0123}=+1$, $D^\mu=\partial^\mu+igA^\mu$ and $\gamma_5=-i\gamma^0\gamma^1\gamma^2\gamma^3$ so that $\bar{u}(PS)\gamma_5\gamma^\mu u(PS)=2S^\mu$. 
 The three GPDs $H^{\pm,\perp}$ are all functions of $x,\Delta^2$ and the skewness parameter $\xi=-\Delta^+/2\bar{P}^+$, with the property $H^*(x,\xi,\Delta^2)=H(x,-\xi,\Delta^2)$. 
 From Lorentz invariance,
 \beq
 \int dx H^+ = \int dx H^- = \int dx H_\perp \equiv H(\Delta^2)
 \eeq
 so that
 \beq
\int dx  \tilde{F}(x,\xi,\Delta^2) =\frac{-i}{2M^2}\langle P'|\tilde{F}^a_{\mu\nu}F_a^{\mu\nu}|P\rangle = \frac{H(\Delta^2)}{2M^2} \bar{u}(P'S')\Slash \Delta \gamma_5u(PS) =  \frac{H(\Delta^2)}{M}\bar{u}(P'S')\gamma_5 u(PS). \label{mom}
 \eeq
 While $\tilde{F}$ vanishes in the forward limit, $H^{\pm,\perp}(x,\Delta)$ is finite in this limit and satisfies $H(x)=H^*(x)=H(-x)$.
 To linear order in $\Delta$, one can approximate 
 \beq
 \bar{u}(P'S')\Slash \Delta \gamma_5u(PS) \approx -2\Delta\cdot S
 \eeq
 if $S'\approx S$. 
 The connection between the operator $F\tilde{F}$ and spin physics is already manifest. 
 
As demonstrated in \cite{Hatta:2020iin},  $F(x)$ in (\ref{ff}) contains the `zero-mode' contribution proportional to $\delta(x)$. On general grounds, one expects that  $\tilde{F}(x)$ also contains the delta function 
 \beq
 H^{\pm,\perp}(x) = H^{\pm,\perp}_{reg}(x) + \delta(x)C^{\pm,\perp}. \label{zero}
 \eeq
 The possible polarization dependence of $C$ has to be canceled by that of the regular part $H_{reg}$ upon integrating over $x$.  
We however conjecture that $C^{\pm,\perp}=0$ based on a prejudice that  spin-dependent distributions are in general suppressed by one power of $x$ compared to spin-independent ones. That is, if $F(x)$ in (\ref{ff}) contains a $\delta(x)$ as shown in \cite{Hatta:2020iin},  naively $\tilde{F}(x)$ does not because $x\delta(x)=0$. In the one-loop calculation in the next section, we shall see an example of this cancellation.

 \subsection{First moment}
 
 Let us study the first moment (\ref{mom}) in more detail. As is well known, the $F\tilde{F}$ operator is related to the flavor-singlet axial current via the $U_A(1)$ anomaly 
 \beq
 \partial_\mu J_5^\mu =2i\sum_f m_f \bar{\psi}_f \gamma_5\psi_f + n_f \frac{\alpha_s}{4\pi} \tilde{F}_{\mu\nu}^a F^a_{\mu\nu}, \label{an}
\eeq
 where $J_5^\mu=\sum_f \bar{\psi}_f\gamma^\mu \gamma_5 \psi_f$ is the  $U_A$(1) current.  
Taking the nonforward matrix element of (\ref{an}), one finds
 \beq
 \langle P'|n_f\frac{\alpha_s}{4\pi}\tilde{F}^a_{\mu\nu}F_a^{\mu\nu}|P\rangle &=& i\Delta_\mu \langle P'|J_5^\mu(0)|P\rangle -2\sum_f \langle P'|m_f\bar{\psi}_fi\gamma_5\psi_f|P\rangle \nn
 &=&i\Delta_\mu \bar{u}(P')\left[\gamma^\mu \gamma^5 F_A(\Delta^2)+\frac{F_P(\Delta^2)}{2M}\Delta^\mu \gamma_5 + i\frac{F_T(\Delta^2)}{2M}\sigma^{\mu\nu}\Delta_\nu \gamma_5 \right]u(P) \nn && \qquad -2\sum_f m_fG_f(\Delta^2)\bar{u}(P')i\gamma_5u(P) \nn
 &=&  2M\left( F_A(\Delta^2)+\frac{\Delta^2}{4M^2}F_P(\Delta^2) -\sum_f \frac{m_f}{M}G_f(\Delta^2)\right)\bar{u}(P')i\gamma_5u(P) \label{chiral}
 \eeq
 where $F_A$, etc. are various form factors.  By definition,
 \beq
 \langle PS|J_5^\mu |PS\rangle =-2S^\mu \Delta \Sigma, \quad \to\quad  F_A(0) = \Delta \Sigma,
 \eeq
 where $\Delta \Sigma$ is the quarks' helicity contribution to the nucleon spin. 
 Since $F_P(\Delta)$ does not have a pole at $\Delta=0$ due to the absence of a massless  singlet pseudscalar meson, one obtains the relation
\beq
\Delta \Sigma-\sum_f\frac{m_f}{M}G_f(0)= \frac{n_f\alpha_s}{4\pi}\int dx H(x).
\eeq

On the other hand, the connection between the anomaly and the gluon helicity contribution $\Delta G$ to the nucleon spin has been extensively discussed in the literature. To see quickly the relevance of $\Delta G$, one introduces  
the topological current 
 \beq
 K^\mu = \epsilon^{\mu\nu\rho\lambda} \left(A^a_\nu F^a_{\rho\lambda} + \frac{g}{3}f_{abc}A_\nu^a A_\rho^b A_\lambda^c \right), \qquad \partial_\mu K^\mu = \tilde{F}^a_{\mu\nu}F_a^{\mu\nu}.
 \eeq 
 In the light-cone gauge $A^+=0$,
 $K^+ = 2\epsilon^{ij}A_i^a \partial^+A_j^a$, and its nucleon matrix element is related to $\Delta G$. 
 \beq
\left. \langle P|K^+|P\rangle \right|_{A^+=0} = 4S^+\Delta G.
 \eeq
However, the other components of $K^\mu$ bear no simple relation to $\Delta G$, nor do they have a well defined forward matrix element \cite{Balitsky:1991te}. 
  Consider then the modified current which is approximately conserved 
 \beq
\tilde{J}_5^\mu \equiv J_5^\mu -n_f\frac{\alpha_s}{4\pi}K^\mu, \qquad  \partial_\mu \tilde{J}^\mu_5=2i\sum_f m_f\bar{\psi}_f\gamma_5\psi_f.
 \eeq
 Notice that
 \beq
\left. \langle P|\tilde{J}^+_5|P\rangle \right|_{A^+=0} = -2S^+\left(\Delta \Sigma + n_f\frac{\alpha_s}{2\pi}\Delta G \right)
\eeq
 Since $\tilde{J}_5^\mu$ is conserved, naively one expects that the linear combination is scale independent 
 \beq
 \frac{\partial}{\partial \ln \mu^2} \left(\Delta \Sigma(\mu^2) + n_f\frac{\alpha_s(\mu^2)}{2\pi}\Delta G(\mu^2) \right) =0. \label{nor}
 \eeq
  However, the current conservation does not automatically imply vanishing anomalous dimension for the components of $\tilde{J}_5^\mu$.\footnote{ Consider the integral 
 \beq
 \partial_+ \int dx^- d^2x_\perp \tilde{J}_5^+  = -\int dx^- d^2x_\perp (\partial_- \tilde{J}_5^- -\partial_\perp \tilde{J}_5^\perp)
 \eeq
 The usual argument is that the right hand side is the integral of a  total derivative, hence it vanishes.  
This means  $\int d^3x \tilde{J}_5^+$ is a conserved charge and cannot be renormalized. 
 However, in the present case, $\tilde{J}_5^-$ contains $K^-$ which is not gauge invariant. In the light-cone gauge, the matrix element of this operator  has a singularity $K^- \sim \frac{1}{n\cdot \partial}j = \frac{1}{\partial_-}j$ \cite{Balitsky:1991te}. }
As shown in \cite{Kaplan:1988ku},  the non-renormalizability of (\ref{nor}) actually boils down to that of the  operator $\alpha_s F\tilde{F}$ at zero momentum transfer, which is believed to be true to all orders due to its topological nature.  See \cite{deFlorian:2019egz} for a recent application of this property.

 In what follows, we avoid dealing with the gauge variant operator $K^\mu$ which has caused a lot of confusion in the literature. To establish a connection between $F\tilde{F}$ and $\Delta G$ in this case, 
we use the equation of motion relations for $\tilde{F}(x)$. It can be derived similarly to Eq.~(29) of \cite{Hatta:2020iin}, except that we now have to keep the surface terms and  use the Bianchi identity $D^\mu \tilde{F}_{\mu\nu}=0$. The result is  
 \beq
 \tilde{F}_{reg}(x,\Delta) &=& \frac{-i\Delta_\mu}{2xM^2} \int \frac{dz^-}{2\pi}e^{ix\bar{P}^+z^-}\langle P'|\tilde{F}^\mu_{\ \nu}(-z^-/2)F^{\nu+}(z^-/2)-F^{\nu+}(-z/2)\tilde{F}^\mu_{\ \nu}(z/2)|P\rangle \nn&& + \frac{i}{2xM^2} \int \frac{dz^-}{2\pi}e^{ixP^+z^-} \int_{-z^-/2}^{z^-/2}d\omega^- \nn 
 && \times\langle P'|\tilde{F}_{\mu\nu}(-z/2) gF^{+\mu}(\omega^-) F^{+\nu}(z^-/2) - F^{+\nu}(-z/2)F^{+\mu}(\omega^-)\tilde{F}_{\mu\nu}(z/2)|P\rangle, \label{last}
 \eeq
  where Wilson lines are understood. 
 Note that only the regular part (i.e., excluding the delta function $\delta(x)$) can be constrained in this method, see \cite{Hatta:2020iin}.  
Since the first line is multiplied by $\Delta$, to linear order in $\Delta$ one can  take the forward limit and use the general decomposition
 \beq
 P^+\int \frac{dz^-}{2\pi}e^{ixP^+z^-}\langle P|\tilde{F}^\mu_{\ \nu}(0)F^{\nu+}(z^-)|P\rangle = ix\Delta G(x) S^+ p^\mu + 2ixG_{3T}(x)P^+S^{\perp \mu}-ixG_4(x) M^2 S^+n^\mu \label{la}
 \eeq
where $P^\mu=p^\mu+\frac{M^2}{2}n^\mu$, $S^\mu=(S\cdot n)p^\mu + S^\mu_\perp + (S\cdot p)n^\mu$ and $p\cdot n=1$. In the last term one can write  $-M^2S^+/P^+=2P^+S^-$ from $P\cdot S=0$. $\Delta G(x)$ is the usual twist-two polarized gluon distribution $\int_0^1dx \Delta G(x)=\Delta G$. 
$G_{3T}(x)$ is the gluonic analog of the $g_T(x)$ distribution function relevant to the transverse polarization. Its properties have been studied in \cite{Ji:1992eu,Hatta:2012jm,Koike:2019zxc}. $G_4(x)$ is the twist-four counterpart of these distributions whose properties are virtually unknown.  

Eqs.~(\ref{last}) and (\ref{la}) clearly show in a gauge invariant manner that in the near-forward limit $\tilde{F}(x)$ and $\Delta G(x)$ are directly related at the density level (in the longitudinally polarized case). It also shows that they differ by the twist-four, three-gluon correlation function $\sim \tilde{F}FF$. 
A similar relation has been 
 derived in  \cite{Balitsky:1991te} in the light-cone gauge.  The present derivation is manifestly gauge invariant and avoids the subtleties of the light-cone gauge such as the boundary condition at infinity. 
Integrating over $x$, we get 
 \beq
 \int dx \tilde{F}_{reg}(x)
 &\approx & \frac{2}{M^2}(\Delta\cdot S)\Delta G \label{c2}\\ &&  +\frac{i}{M^2}\int dx  \int \frac{dz^-}{2\pi}\frac{e^{ixP^+z^-}}{x} \int_{-\frac{z^-}{2}}^{\frac{z^-}{2}}d\omega^- \langle P'|\tilde{F}_{\mu\nu}(\frac{-z^-}{2}) gF^{+\mu}(\omega^-) F^{+\nu}(\frac{z^-}{2})|P\rangle,   \nonumber
 \eeq
 where  $\int dx \Delta G(x) = 2\int dx G_{3T}(x)=2\int dx G_4(x)=2\Delta G$  from Lorentz invariance.  
 To simplify the notation, 
 let us define
 \beq
 N(x_1,x_2) = \lim_{\Delta\to 0} \frac{1}{\Delta \cdot S} \int \frac{dz^-}{2\pi}\frac{d\omega^-}{2\pi} e^{\frac{i}{2}(x_1+x_2)P^+z^-+i(x_2-x_1)P^+\omega^-} \nn \times  \langle P'|\tilde{F}_{\mu\nu}(-z/2) gF^{+\mu}(\omega^-) F^{+\nu}(z^-/2) |P\rangle  \label{ns}
 \eeq
From $PT$ symmetry,  $N(x_1,x_2)=N(x_2,x_1)$. 
Using this and  equating  (\ref{chiral}) with  (\ref{c2}) in the near-forward limit, we arrive at 
  \beq
  \Delta \Sigma + \frac{n_f\alpha_s}{2\pi} \Delta G - \sum_f \frac{m_f}{M}G_f(0) -\frac{n_f\alpha_s}{4\pi}C &=&\frac{n_f\alpha_s}{2\pi}\int dx_1dx_2{\cal P}\frac{N(x_1,x_2)}{x_1(x_1-x_2)} \nn &=& -\frac{n_f\alpha_s}{4\pi}\int dx_1dx_2\frac{N(x_1,x_2)}{x_1x_2}.
  \label{nob}
 \eeq
 In this formula (actually already in (\ref{ns})),  we have  assumed  that $C$ does not depend on the spin orientation $\pm,\perp$. If this turns out not to be the case, the formula must be modified accordingly. (As already mentioned, we suspect that $C=0$ anyway.) 
  
   Among the various terms in (\ref{nob}), 
 the $u,d$-quark mass contributions  can be safely neglected because $m_{u,d}/M< 0.01$ and $G_{u,d}$ is naturally order unity. The impact of the $s$-quark $m_s/M\sim 0.1$ might not be negligible, though. The value of $G_{s}(0)$ can be studied in lattice QCD for instance.  Eq.~(\ref{nob}) shows  that the  RG-invariant linear combination of the twist-two quantities $\Delta \Sigma$ and $\Delta G$ is directly  related to the nonforward matrix element of a twist-four, three-gluon correlator. According to the previous discussion, the latter has to be scale invariant.  To our knowledge (\ref{nob}) has not been presented in this explicit form in the literature, although Ref.~\cite{Balitsky:1991te} comes close.

 \section{One-loop calculations}
 
 In this section, we compute  $\tilde{F}(x)$ for quark and gluon targets in perturbation theory to one-loop in dimensional regularization in $d=4-2\epsilon$ dimensions. We shall only focus on the divergent part to investigate the UV structure of the distribution. Calculating the finite part should also be possible, but  this involves  extra complications regarding the definition of  $\epsilon^{\mu\nu\rho\lambda}$ in $d\neq 4$ dimensions. 
 
 \subsection{Quark target}
We work in the light-cone gauge $n\cdot A=A^+=0$ to eliminate the Wilson line. For an on-shell quark target $(p\pm \Delta/2)^2=m^2$ with $p\cdot \Delta=0$, a straightforward calculation gives,  to linear order in $\Delta$,
 \beq
 \tilde{F}(x) 
 =i\frac{4g^2C_Fp^+}{m^2}\int \frac{dk^-d^{d-2}k_\perp}{(2\pi)^d} 
  \frac{k\cdot \Delta  k\cdot S-k^2\Delta\cdot S }{ (k^2+i\epsilon)^2} \left(\frac{1} {(p-k)^2-m^2+i\epsilon}+\frac{1}{(p+k)^2-m^2+\epsilon}\right),
 \eeq
 where $C_F=\frac{N_c^2-1}{2N_c}$ and $n\cdot k=x$. The first and second terms are nonzero for $1>x>0$ and  $0>x>-1$, respectively. 
 The $dk^- d^2k_\perp$ integral cannot be done all at once. Different components of $k\cdot \Delta k\cdot S=k^\mu k^\nu \Delta_\mu S_\nu$ have to be  evaluated separately.
 The most nontrivial integral is 
 \beq
 I= \int \frac{dk^- d^{d-2}k_\perp}{(2\pi)^{d-1}} \frac{(k^-)^2}{(k^2)^2((p-k)^2-m^2)}
 \eeq
 To evaluate this we write $(k^-)^2=\left(\frac{k^2+k_\perp^2}{2k^+}\right)^2$ and  cancel some denominators. We then use the formula 
 \beq
\int \frac{dk^-}{2\pi} (k^+k^--M^2+i\epsilon)^{-\epsilon}= i\frac{(-1)^\epsilon}{\epsilon-1}\delta(k^+)(M^2)^{1-\epsilon},
\eeq
to get 
\beq 
I
=i\frac{\Gamma(\epsilon)}{4\pi} \frac{p^-}{4(p^+)^2} \left(\delta(1-x)-2x\right) .
\eeq
The other integrals are straightforward to evaluate. 
The result is,  for $1>x>0$, 
\beq
\tilde{F}(x)=\frac{\alpha_sC_F\Gamma(\epsilon)}{2\pi m^2} \left((2-x)\Delta^-S^+ + (2+\delta(1-x)-3x)\Delta^+S^- -(1+x)\vec{\Delta}_\perp \cdot \vec{S}_\perp \right).
\eeq
Comparing this with (\ref{last}) and (\ref{la}), we obtain 
\beq
\Delta G(x) &=& \frac{\alpha_sC_F\Gamma(\epsilon)}{2\pi} (2-x),\nn 
G_{3T}(x) &=& \frac{\alpha_sC_F\Gamma(\epsilon)}{2\pi}\frac{1+x}{2}, \nn 
G_4(x)&=& \frac{\alpha_sC_F\Gamma(\epsilon)}{2\pi} \frac{2+\delta(1-x)-3x}{2}.
\eeq
This identification is possible because the twist-four correlator $\langle g\tilde{F}FF\rangle$ is at least ${\cal O}(\alpha_s^2)$ for a quark target.  
One immediately recognizes the polarized splitting function $\Delta P_{gq}(x)=C_F(2-x)$ in the longitudinal sector. 
Note that the delta function at $x=0$ is absent. 
 Once integrated over $x$, $\tilde{F}$ becomes proportional to $\Delta \cdot S$ as it should
 \beq
\int_{-1}^1 dx \tilde{F}(x)= \frac{3\alpha_sC_F\Gamma(\epsilon)}{2\pi m^2} \Delta \cdot S.
 \eeq
This leads to
\beq
\Delta \Sigma -\frac{m}{M}G_F(0)= -6n_f C_F \left(\frac{\alpha_s}{4\pi}\right)^2 \Gamma(\epsilon),
\eeq
which reproduces the known anomalous dimension of the axial current operator \cite{Adler:1969gk,Kodaira:1979pa}
\beq
\gamma = -6n_f C_F\left(\frac{\alpha_s}{4\pi}\right)^2.
\eeq



\subsection{Gluon target}
\label{gluon}

For regularization purpose, we assume that the incoming and outgoing gluons are slightly off-shell $(p- \Delta/2)^2=(p+\Delta/2)^2<0$. The initial and final polarization vectors $\varepsilon_{i/f}$ satisfy $\varepsilon_i\cdot p=\varepsilon_i \cdot\Delta/2$ and $\varepsilon_f \cdot p = -\varepsilon_f \cdot \Delta/2$, respectively, and the ${\cal O}(\Delta)$ terms must be kept. The diagrams to be calculated are identical to those in the case of the $FF$ correlator  \cite{Hatta:2020iin}, but the off-forward kinematics brings in  considerable complications. For simplicity, in the following  we assume $\Delta^+=0$. This approximation significantly  reduces the number of terms in intermediate calculations while keeping the most important term $\sim \Delta^-S^+$ relevant to longitudinal polarization.  After a very tedious calculation, the sum of the connected  diagrams (i.e., without the self-energy diagrams on external legs) is found to be, for $1>x>0$, 
\beq
M^2\tilde{F}(x)&=& \delta(1-x) i\epsilon^{\Delta  \varepsilon_f^*\varepsilon_i p}+ \frac{g^2N_c}{2}  \Biggl[ \int \frac{dk^- d^2k_\perp}{(2\pi)^4} \frac{p^+I_1}{x(1-x)(k^2+i\epsilon)^2(p-k+i\epsilon)^2}\nn
&&+\delta(1-x) \int \frac{d^4k}{(2\pi)^4}\frac{(p^+)^2I_2}{k^+(p^+-k^+)(k^2+i\epsilon)^2(p-k+i\epsilon)^2} \Biggr]\epsilon^{\Delta  \varepsilon_f^*\varepsilon_i p},  
\eeq
 where $\epsilon^{ABCD}\equiv \epsilon_{\mu\nu\rho\lambda}A^\mu B^\nu C^\rho D^\lambda$ and 
 \beq
 I_1 
= 12x(x-1)k_\perp^2 -6x^2(1-x)(k-p)^2 +4x^2(3-2x)k^2 + 4x^2 (2x-1)(x-1)p^2.
\eeq
 \beq
I_2=
 2x(1-x)(p-k)^2 +2x(x-1-2x^2)p^2 + 2x(2x-1)k^2+2(x-1)k_\perp^2.
\eeq
At this point we may set $\varepsilon_i=\varepsilon_f$ and drop the subscripts $i/f$. 
To arrive at the above result, we used the following relations which hold only when $\Delta^+=0$ \beq
&&\epsilon^{\Delta \varepsilon^* \varepsilon n}=0, \qquad \epsilon^{\Delta \varepsilon^* \varepsilon k}=x\epsilon^{\Delta \varepsilon^* \varepsilon p}, \nn 
&& \epsilon^{\varepsilon^*\varepsilon kn}\Delta \cdot k= 
-x^2\epsilon^{\Delta \varepsilon^* \varepsilon p}, \qquad \Delta \cdot k \epsilon^{\varepsilon^*\varepsilon kp}=-\frac{1}{2}(x^2p^2-k^2-k_\perp^2) \epsilon^{\Delta \varepsilon^* \varepsilon p}
\eeq
 To proceed, following \cite{Hatta:2020iin}, we employ the Mandelstam-Leibbrandt (ML) prescription for the spurious poles in the light-cone gauge
 \beq
 \frac{1}{[k^+]_{\rm ML}} = \frac{1}{k^+ + i\epsilon k^-},\qquad \frac{1}{[p^+-k^+]_{\rm ML}} = \frac{1}{p^+ - k^+ + i\epsilon (p^- - k^-)}.
 \eeq
 The remaining integrals can  be done using the formulas collected in an appendix of \cite{Hatta:2020iin} and other formulas such as
 \beq
 \int \frac{d^4k}{(2\pi)^4} \frac{p^2}{[p^+-k^+]_{\rm ML}(k^2)^2 (p-k)^2} = \frac{i\Gamma(\epsilon)}{16\pi^2 p^+}, \qquad
\int \frac{d^4k}{(2\pi)^4} \frac{k_\perp^2}{[k^+]_{\rm ML}(k^2)^2(p-k)^2}= \frac{4i\Gamma(\epsilon)}{16\pi^2 p^+} 
\eeq
 Including also the self-energy diagrams, our final result is, for $1>x>0$, 
 \beq
 M^2\tilde{F}(x)= \Delta\cdot S \left[ \delta(1-x)  + \Gamma(\epsilon) \frac{\alpha_s}{2\pi} \left( \Delta P_{gg}(x) +2N_c\left(1+x-\frac{3}{2}\delta(1-x)\right) \right)  \right],
 \eeq
  where 
  \beq
  S^\mu = i\epsilon^{\mu \varepsilon^* \varepsilon p},
  \eeq
  is the spin four-vector for a spin-1 particle.  In the ${\cal O}(\alpha_s)$ terms, we have separated out the polarized splitting function
  \beq
  \Delta P_{gg}(x) = 2N_c \left(1-2x+\frac{1}{[1-x]_+}\right) + \frac{\beta_0}{2}\delta(1-x),
  \eeq
where $\beta_0=\frac{11N_c}{3} -\frac{2n_f}{3}$. The remainder terms
 \beq
 \Delta \tilde{P}(x)= 2N_c\left(1+x-\frac{3}{2}\delta(1-x)\right),
 \eeq
  come from the twist-four operator $g\tilde{F}FF$ which has nonvanishing gluon matrix element to ${\cal O}(\alpha_s)$. A useful consistency check is that the integral has to vanish $\int_0^1 dx \Delta \tilde{P}(x)=0$.  This guarantees that the renormalization of the local operator $F\tilde{F}$ is entirely due to the charge renormalization. In other words,   $\alpha_s F\tilde{F}$ is renormalization-group invariant.    
 Note that  again there is no delta function $\delta(x)$. Interestingly, the second term of $I_1$ potentially gives rise to a delta function from the integral
 \beq
 \int dk^- d^2k_\perp \frac{1}{(k^2+i\epsilon)^2} \propto \delta(x).
 \eeq
 However, there remains one factor of $x$ in the numerator which kills this delta function $x\delta(x)=0$, see a similar example in \cite{Burkardt:2001iy}. This is consistent with our previous claim that $C$ might actually be zero.

 \section{Third moment}
 
 This section is to a large extent inspired by the work of Seng  \cite{Seng:2018wwp} which tried to establish a link between  higher-twist parton  distributions  
 and the so-called quark chromo-magnetic dipole moment operator
 \beq
 \langle P|\bar{\psi}gF^{\mu\nu}\sigma_{\mu\nu}\psi|P\rangle. \label{se}
 \eeq 
 This matrix element is important in the context of $CP$-violation in low energy hadron physics, in particular,  the electric dipole moment (EDM) of the nucleons. 
 While the operator (\ref{se})  itself does not violate $CP$, via chiral symmetry its matrix element is   proportional to $CP$-violating effective low energy interactions (see, e.g.,  \cite{deVries:2016jox}). The idea of \cite{Seng:2018wwp} is that one can get information about this  matrix element from the  chiral-odd twist-three distribution
 \beq
 e(x) =\frac{P^+}{2M}\int \frac{dz^-}{2\pi} e^{ixP^+z^-} \langle P|\bar{\psi}(0)W \psi(z^-)|P\rangle,
 \eeq
accessible in high energy reactions such as semi-inclusive DIS (SIDIS) \cite{Jaffe:1991ra,Balitsky:1996uh,Efremov:2002ut}.

 Specifically, the third moment of $e(x)$ reads 
 \beq
 \int dx x^2 e(x) = \frac{1}{4M(P^+)^2} \langle P|\bar{\psi}g F^{+\mu}\sigma^+_{\ \mu}\psi|P\rangle + \cdots
 \label{se2}
 \eeq
 where the neglected terms are relatively better under control. 
 The operators in (\ref{se}) and (\ref{se2}) indeed look similar, but they are crucially different in the way Lorentz indices are treated. In other words, they have different twists, and the matrix elements of operators with different twists are in general unrelated, unless one makes  extra assumptions as was done in \cite{Seng:2018wwp}. While the validity of such  assumptions must be carefully scrutinized, that is not the purpose of this paper. Here instead, we point out an analogous, tantalizing  connection between the third moment of $\tilde{F}(x)$ and the matrix element of the so-called Weinberg operator \cite{Weinberg:1989dx}
 \beq
 {\cal O}_W=g f_{abc}\tilde{F}_{\mu\nu}^a F_b^{\mu\alpha}F_{c\alpha}^\nu.
 \eeq
 This operator violates $CP$  and can be induced in the QCD Lagrangian by physics beyond the Standard Model. It is considered as one of the candidate operators to generate a large electric dipole moment (EDM)  of the nucleons and nuclei. 
 
At a superficial level, the connection can be readily seen by computing the third moment 
 \beq
 I_3 \equiv \int dx x^2 \tilde{F}(x) &=& \frac{i}{2M^2(\bar{P}^+)^2}\langle P'|\tilde{F}_{\mu\nu}(\overleftrightarrow{D}^+)^2F^{\mu\nu}(0)|P\rangle \nn 
 &=&\frac{1}{M^2(\bar{P}^+)^2}\Delta^\mu \langle P'|\tilde{F}_{\mu\nu} \overleftrightarrow{D^+}F^{\nu+}(0)|P\rangle  - \frac{1}{(\bar{P}^+)^2M^2} \langle P'|\tilde{F}_{\mu\nu}(0) gF^{+\mu}(0)F^{+\nu}(0)|P\rangle ,
   \label{gen}
 \eeq
where  $\overleftrightarrow{D}^+\equiv \frac{D^+ - \overleftarrow{D}^+}{2}$. The three-gluon operator on the second line is similar to the Weinberg operator, but it has open Lorentz indices $++$ as a remnant of the underlying light-cone distribution.  This is entirely analogous to the difference between (\ref{se}) and (\ref{se2}).  
To better appreciate this difference, let us consider the various matrix elements in (\ref{gen}) in more detail. In fact, (\ref{gen}) is a special case of the following more general operator identity\footnote{ To prove (\ref{ne}), the following identity is useful
\beq
[D_\beta,[D_\nu, F_{\alpha\mu}]]^a -[D_\nu,[D_\beta, F_{\alpha\mu}]]^a=gf_{abc}F^b_{\alpha\mu}F^c_{\beta\nu}. 
\eeq 
}
\beq
f^{abc}\tilde{F}^a_{\mu\nu} gF_b^{\alpha\mu}F_c^{\beta\nu}&=& 
-\partial^\mu(\tilde{F}_{\mu\nu} \overleftrightarrow{D}^{(\beta} F^{\nu\alpha)}) -\frac{1}{2}\tilde{F}_{\mu\nu}\overleftrightarrow{D}^{(\beta} \overleftrightarrow{D}^{\alpha)} F^{\mu\nu} \nn 
&=& 
-\partial^\mu(\tilde{F}_{\mu\nu} D^{(\beta} F^{\nu\alpha)}) -\frac{1}{2}\tilde{F}_{\mu\nu}D^{(\beta} D^{\alpha)} F^{\mu\nu} \nn 
&=&-\partial^\mu(\tilde{F}_{\mu\nu}D^\beta F^{\nu\alpha})-\frac{1}{2}\tilde{F}_{\mu\nu}D^\beta D^\alpha F^{\mu\nu} \label{ne}
\eeq
where $(\alpha\beta)$ denotes symmetrization of indices, e.g., $A^{(\alpha}B^{\beta)}=\frac{A^\alpha B^\beta + A^\beta B^\alpha}{2}$.
The matrix element of the total derivative operator on the right hand side of (\ref{ne})  can be  essentially determined by  observables in  QCD spin physics. Since this is multiplied by $\partial^\mu \sim \Delta^\mu$, it is enough to consider the forward matrix element
\beq
\langle PS| \tilde{F}^\mu_{\ \nu}D^{(\alpha} F^{\nu \beta)}|PS\rangle &=&\frac{2a_2}{3} \left(S^\mu P^\alpha P^\beta +(S^\alpha P^\beta + S^\beta P^\alpha)P^\mu -\frac{M^2}{6}(g^{\alpha\beta}S^\mu+g^{\alpha\mu}S^\beta + g^{\beta\mu}S^\alpha)  \right) 
\nn
&&+ \frac{2d_2}{3}\left( 2S^\mu P^\alpha P^\beta-(S^\alpha  P^\beta+S^\beta  P^\alpha)P^\mu   + \frac{M^2}{3}(-2g^{\alpha\beta}S^\mu + g^{\alpha\mu}S^\beta + g^{\beta\mu}S^\alpha) \right) \nn 
&& +\frac{f_0 M^2}{18} ( 5 g^{\alpha\beta} S^\mu -g^{\alpha\mu}S^\beta -g^{\beta\mu}S^\alpha) \label{t2}
\eeq
where the first, second and third lines correspond to twist-2,3,4 parts of the operator, respectively. To get this structure note that $S\cdot P=0$ and require that the tensor vanishes after summing over  $\mu$ and $\alpha$ (or $\mu$ and $\beta$) because  $\tilde{F}_{\mu\nu} ( D^\mu F^{\nu\beta}+D^\beta F^{\nu\mu}) =\partial^\mu(\tilde{F}_{\mu\nu}F^{\nu\beta}) +\frac{1}{2} \partial^\beta (\tilde{F}_{\mu\nu}F^{\nu\mu})$ is a total derivative operator. 
In particular, the trace part reads
\beq
\langle PS| \tilde{F}^\mu_{\ \nu}D_\alpha F^{\nu \alpha}|PS\rangle
=
-\langle P|\bar{\psi}g\tilde{F}^{\mu\nu}\gamma_\nu \psi|P\rangle =f_0 M^2 S^\mu \label{4}
\eeq
 where we used the equation of motion. The parameter $f_0$ shows up as part of the twist-four corrections to the first moment of the $g_1$ structure function  in polarized DIS   \cite{Shuryak:1981pi,Balitsky:1989jb,Ji:1993sv,Kawamura:1996gg}.
On the other hand, $a_2$ and $d_2$ are related to the third moment of $\Delta G(x)$ and $G_{3T}(x)$ as \cite{Hatta:2012jm}
\beq
\frac{1}{2} \int dx x^2 \Delta G(x) =a_2, \qquad  \int dx x^2 G_{3T}(x) = \frac{a_2+2d_2}{3}.
\eeq
 Thus, at least in principle, these parameters can be constrained by  high energy polarized hadron collision experiments.

Next consider the matrix element of the three-gluon operator on the left hand side of (\ref{ne}). Its general parametrization is   
 \beq
&& \frac{1}{M^2}\langle P'|gf^{abc}\tilde{F}_{\mu\nu}^a F^{\alpha\mu}_bF^{\beta \nu}_c|P\rangle \nn
&& = \bar{u}(P')\Biggl[A \left(\Delta^{(\alpha} \gamma^{\beta)} -\frac{\Slash \Delta g^{\alpha\beta}}{4}\right)i\gamma_5 + \frac{B}{M} \left(P^\alpha P^\beta-\frac{g^{\alpha\beta}M^2}{4}\right) i\gamma_5  
 +\frac{C}{M}\left(\Delta^\alpha\Delta^\beta-\frac{\Delta^2g^{\alpha\beta}}{4}\right)
i\gamma_5  +g^{\alpha\beta}i\gamma_5  E M \Biggr]u(P) \nn 
&&\approx -2iA(0)\left(\Delta^{(\alpha} S^{\beta)} -\frac{\Delta \cdot S g^{\alpha\beta}}{4}\right)  -i\frac{B(0)}{M^2}  \left(P^\alpha P^\beta -\frac{g^{\alpha\beta}M^2}{4}\right)\Delta \cdot S -ig^{\alpha\beta}E(0)\Delta \cdot S  \label{zo}
 \eeq
 where $A,B,C,E$ are dimensionless form factors (all functions of $\Delta^2$).
 In the last line of (\ref{zo}) we took the limit $\Delta\to 0$ and kept only the terms linear in $\Delta$. The Weinberg operator is related to the $E$ form factor
\beq
\frac{1}{M^2}\langle P'|gf^{abc}\tilde{F}_{\mu\nu}^a F_b^{\mu\sigma} F_{c\sigma}^\nu|P\rangle = 
 4E(\Delta^2) M \bar{u}(P')i\gamma_5 u(P).
\label{wei}
\eeq
Plugging the $\alpha\beta=++$ component of (\ref{t2}) and (\ref{zo}) into (\ref{gen}), we find
\beq
I_3= \frac{\Delta \cdot S}{M^2}\left(2\int dx x^2G_{3T}(x) -B(0)\right) + \frac{\Delta^+S^-}{M^2}\left(4A(0) +\frac{2}{9}(2a_2-4d_2+f_0)\right). \label{ii3}
\eeq
Note that the second term is absent if one considers  transverse polarization $S^\mu = \delta^\mu_i S^i_\perp$. 

Eq.~(\ref{ii3}) is as far as one can get based only on general principles such as symmetries and the equation of motion. 
It confirms our previous expectation  that there is in general no relation between the third moment $I_3$ and the matrix element of the Weinberg operator $E(0)$. However, there may be hidden relations among different form factors which follow from the dynamics of the theory. For example, if one  naively (perhaps unjustifiably) applies the argument of \cite{Seng:2018wwp} to the present problem, one   finds $E(0)\sim -\frac{3B}{4}$ and  $I_3$ becomes sensitive to $E(0)$.

 \section{Connection between the Weinberg operator and polarized DIS}
 
Quite independently of `hidden relations' just mentioned, our analysis in the previous section has revealed an important, model-independent feature of the Weinberg operator. Taking  the matrix element of the 
the trace of (\ref{ne})
\beq
{\cal O}_W &=&  -\partial^\mu (\tilde{F}_{\mu\nu} \overleftrightarrow{D}_\alpha F^{\nu\alpha}) -\frac{1}{2} \tilde{F}_{\mu\nu}\overleftrightarrow{D}^2 F^{\mu\nu} \nn 
&\equiv& {\cal O}_4 + {\cal O}_D, \label{ind}
\eeq
we find
\beq
E(0)=\frac{f_0}{4} + \frac{1}{8iM^2}\lim_{\Delta\to 0} \frac{1}{\Delta\cdot S}\langle P'|\tilde{F}_{\mu\nu}(\overleftrightarrow{D})^2F^{\mu\nu}|P\rangle. \label{compa}
\eeq
This shows that $E(0)$ is related to the parameter $f_0$ that enters the twist-four corrections to the $g_1$ structure function,  unless $f_0$ is completely canceled by the unknown matrix element $\sim \langle \tilde{F}D^2F\rangle$. We can actually exclude the latter possibility using the following renormalization group (RG) argument.  Eq.~(\ref{ind})  shows that one can choose ${\cal O}_W$ and ${\cal O}_4$ as the independent basis of operators and study their mixing.\footnote{Using the identities $D_\mu\tilde{F}^{\mu\nu}=0$ and $D_{\mu}F_{\alpha\beta}+D_\alpha F_{\beta\mu}+D_\beta F_{\mu\alpha}=0$, one sees that there are no other independent, pseudoscalar, dimension-six gluonic operators up to one total derivative. 
 We  also neglect the mixing with  the  quark chromo-electric dipole moment  operator $m\bar{\psi}gF_{\mu\nu}\sigma^{\mu\nu}\gamma_5 \psi$ \cite{Morozov:1985ef,Morozov:1983qr,Braaten:1990gq} assuming massless quarks.   } To linear order in $\partial\sim \Delta$, this is equivalent to considering  the operator 
\beq
{\cal O}_4 \approx \partial^\mu (\bar{\psi}g\tilde{F}_{\mu\nu}\gamma^\nu\psi),
\eeq
 due to the equation of motion.  Such mixing is usually neglected in the literature  because ${\cal O}_4$ is a total derivative and hence does not contribute to the $CP$-violating effective action $\int d^4x {\cal O}_4=0$. However, when it comes to hadronic matrix elements,  mixing becomes crucial because only the nonforward matrix element is nonvanishing.\footnote{For general discussions of mixing with total derivative operators, see \cite{Gracey:2009da,Braun:2011dg}.
} 
Specifically, their RG equation takes the form
 \beq
 \frac{d}{d\ln \mu^2} \begin{pmatrix} {\cal O}_W \\ 
 {\cal O}_4 \end{pmatrix} = -\frac{\alpha_s}{4\pi} 
 \begin{pmatrix} \gamma_W & \gamma_{12} \\ 0  & \gamma_4 \end{pmatrix} \begin{pmatrix} {\cal O}_W \\ 
 {\cal O}_4 \end{pmatrix} \label{mat}
 \eeq
 where 
  \cite{Morozov:1985ef,Morozov:1983qr,Braaten:1990gq}
\beq
\gamma_W= \frac{N_c}{2}+n_f+\frac{\beta_0}{2}  =  \frac{7}{3}N_c +\frac{2}{3}n_f
\eeq
[The factor $\beta_0/2$ comes from the explicit QCD coupling $g$ multiplying the operator in our convention.]
The anomalous dimension of ${\cal O}_4$ is the same as that of the undifferentiated, twist-four operator  $\bar{\psi}g\tilde{F}^{\mu\nu}\gamma_\nu \psi$ 
and is known to be \cite{Shuryak:1981kj,Morozov:1983qr}
\beq
\gamma_4 = \frac{8}{3}C_F+\frac{2}{3}n_f . \label{ga4}
\eeq

\begin{figure}
  \includegraphics[angle=270,width=0.8\linewidth]{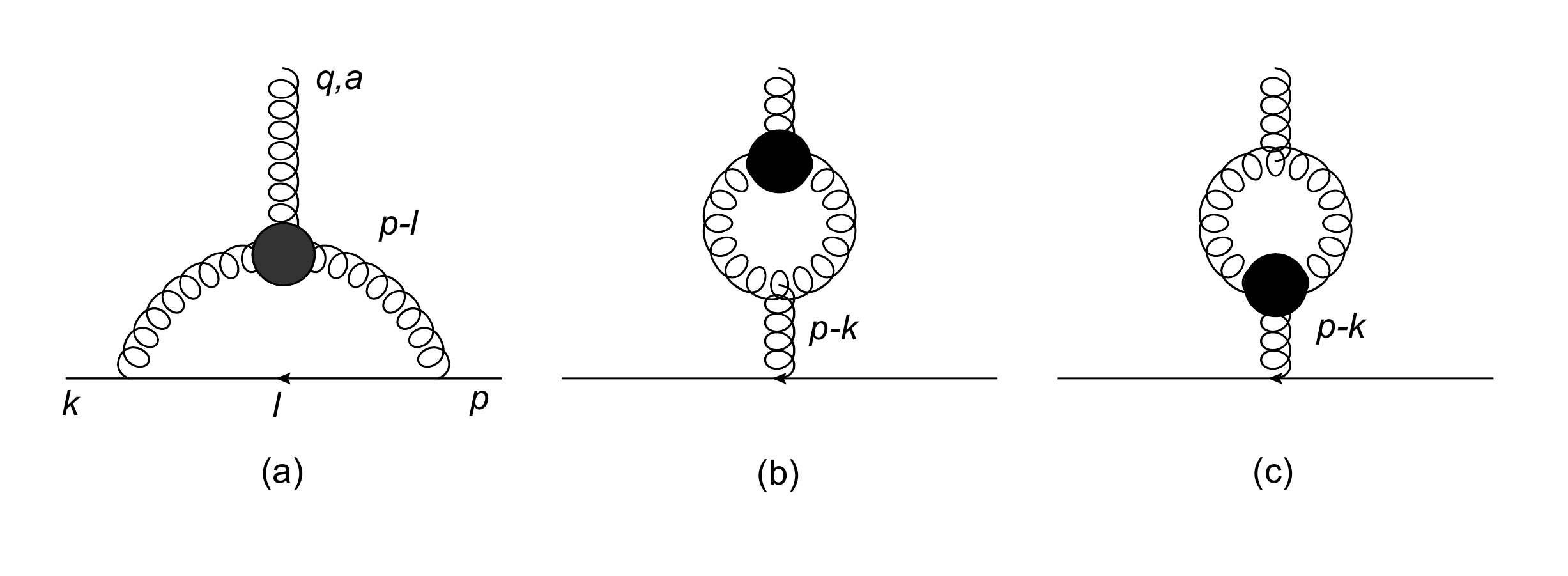}
\caption{Diagrams that contribute to the mixing between ${\cal O}_W$ and ${\cal O}_4$. The black dot denotes the insertion of  ${\cal O}_W$. 
}
\label{fig}
\end{figure}

To determine the off-diagonal component $\gamma_{12}$, we evaluate the following three-point Green's function 
\beq
\langle 0|{\rm T}\{\psi(-k)A_a^\rho(q)\bar{\psi}(p) {\cal O}_W\}|0\rangle
\eeq
with off-shell momenta and  nonzero momentum transfer $\Delta=k-p-q \neq 0$. There are three diagrams as shown in Fig.~\ref{fig}. It is convenient to use the compact Feynman rules suggested in   \cite{Braaten:1990gq}.\footnote{The normalization of ${\cal O}_W$ in \cite{Braaten:1990gq} differs from ours by a factor $-3g$. Also, the sign convention of $\gamma_5$ is opposite to ours (but   $\epsilon^{0123}=+1$ is the same).  } The first diagram gives
\beq
ig^2f^{abc}t^ct^b\int \frac{d^d\ell}{(2\pi)^d} \frac{\gamma^\nu  \Slash \ell\gamma^\mu}{(p-\ell)^2 \ell^2 (\ell-k)^2} \frac{-3g}{16}{\rm Tr}\bigl[[\Slash q,\gamma^\rho][\Slash p -\Slash \ell,\gamma^\mu][\Slash \ell -\Slash k,\gamma^\nu ]\gamma_5\bigr]. \label{kao}
\eeq
Since the gamma matrix  trace provides the necessary  antisymmetric tensor, we may replace\footnote{One can check that the neglected term  $\sim \epsilon^{\mu\nu\rho\lambda}\gamma_\rho \gamma_5\ell_\lambda$ in (\ref{neg})  vanishes after the $\ell$-integral.}
\beq
\gamma^\nu  \Slash \ell \gamma^\mu \to \ell^\nu \gamma^\mu + \ell^\mu \gamma^\nu -g^{\mu\nu}\Slash \ell. \label{neg}
\eeq
 and find
\beq
{\rm diagram \ (a)}=-\frac{3N_cg^3}{16\pi^2}\Gamma(\epsilon)gt^a \gamma_\mu (-\epsilon^{\mu\rho kq}+\epsilon^{\mu\rho pq}) = \frac{3N_c \alpha_s}{4\pi}\Gamma(\epsilon) gt^a \gamma_\mu\epsilon^{\mu\rho \Delta q}.
\eeq
The second diagram is `one-particle reducible' (1PR) \cite{Kodaira:1997ig} and contains the propagator pole $1/(p-k)^2$. After the loop integral, the numerator becomes proportional to $(p-k)^2$  as well as to $\Delta$, so the pole disappears. The result is
\beq
{\rm diagram\  (b)}=-\frac{3N_c \alpha_s}{4\pi}\Gamma(\epsilon) gt^a \gamma_\mu\epsilon^{\mu\rho \Delta q}, \label{cancel}
\eeq
 which cancels the first diagram. 
The third diagram also contains $1/(p-k)^2=1/(q+\Delta)^2$, while the numerator is proportional to $q^2$ (as well as $\Delta$). To linear order in $\Delta$, one can approximate $q^2/(q+\Delta)^2\approx 1$ and find the same result (\ref{cancel}). 
Finally, the tree-level matrix element of ${\cal O}_4$ is 
\beq
\langle {\rm T}\{ \psi(-k)A_a^\rho(q)\bar{\psi}(p) {\cal O}_4\}\rangle  =-gt^a\gamma_\mu \epsilon^{\mu\rho \Delta q}
\eeq
From these results, we deduce that 
\beq
\gamma_{12}=-3N_c. \label{app}
\eeq

It immediately follows that the following linear combination is the  eigenstate of the RG evolution 
\beq
{\cal O}_W +\frac{\gamma_{12}}{\gamma_W - \gamma_4}{\cal O}_4 ={\cal O}_W - \frac{9N_c^2}{3N_c^2+4}{\cal O}_4.
\eeq
Since this operator has a rather large anomalous dimension $\gamma_W\sim 10$, in particular larger than $\gamma_4$ by  a factor of about 2, at  high enough renormalization scales $\mu^2$ one has 
\beq
\langle {\cal O}_W \rangle \approx \frac{9N_c^2}{3N_c^2+4} \langle {\cal O}_4 \rangle \approx 2.61\langle {\cal O}_4\rangle,
\eeq
or equivalently, 
\beq
E \approx \frac{9N_c^2}{4(3N_c^2+4)}f_0 \approx 0.65 f_0.
\eeq
Comparing with (\ref{compa}), we see that the operator $\tilde{F}D^2F$ also contributes to the trace part.

We have thus argued that the matrix element of the Weinberg operator is dominated by its mixing with the total derivative operator ${\cal O}_4$ which is further related to the twist-four operator relevant to polarized DIS.  
Our result urges  one to revisit  previous estimates of $\langle {\cal O}_W\rangle$. For instance, Ref.~\cite{Bigi:1990kz} suggested the following  ansatz
\beq
\langle N|\frac{g^3}{16\pi^2} f^{abc}F_{\mu\nu}^a F^{\mu\alpha}_b F_{c\alpha}^\nu|N\rangle 
=\Lambda_{QCD}^2 \langle N|\frac{\alpha_s}{4\pi} F_{\mu\nu}^a F^{\mu\nu}_a|N\rangle.
\eeq
While such a relation may give a reasonable order-of-magnitude estimate, it has to be interpreted with  great care. Both sides vanish in the forward limit and in the off-forward case the matrix elements are sensitive to the spin polarization. If one tries to relate the coefficients of $\Delta \cdot S$ in the near-forward limit, the right hand side essentially gives  $\Delta \Sigma$, the quark helicity contribution to the nucleon spin,  while the left hand side is related to the parameter $f_0$ which enters the twist-four corrections in polarized DIS as we have shown. There is no known relation between the two quantities.


\section{Conclusions} 

In this paper we have studied the roles of $CP$-odd gluonic operators $\tilde{F}^{\mu\nu}F_{\mu\nu}$ and  $\tilde{F}_{\mu\nu}F^{\mu\alpha}F^{\nu}_{\alpha}$ in  QCD spin physics. These high-dimension, high-twist operators usually do not appear in the standard description of spin-dependent phenomena in terms of twist-two (and sometimes twist-three) distributions.  However, with the future Electron-Ion Collider poised to reveal the gluonic contributions to the nucleon spin and various polarization observables, it is worthwhile and maybe necessary to expand our scope to the twist-four sector. Indeed, we have shown in (\ref{nob}) that the twist-two observables $\Delta\Sigma$ and $\Delta G$ are related to a certain twist-four correlator. Moreover,  $\tilde{F}(x)$ directly shows up in a recent calculation of the $g_1(x)$ structure function \cite{Tarasov:2020cwl}. As we have seen, $\tilde{F}(x)$ contains $\Delta G(x)$, and this should be taken into account when fully extracting the implications of the result in  \cite{Tarasov:2020cwl}.  Concerning the dimension-six, Weinberg operator $\tilde{F}_{\mu\nu}F^{\mu\alpha}F^{\nu}_{\alpha}$, hopefully our result better motivates a precise determination of the parameter $f_0$ through the measurement of the $g_1(x)$ structure function. This could be a useful input to the studies of the nucleon electric dipole moment.

\acknowledgments
I am grateful to the Yukawa Institute for Theoretical Physics at Kyoto university for hospitality during the whole period of this work.
I thank Kazuhiro Tanaka and Yong Zhao for discussions and Raju Venugopalan for correspondence.  This work is supported by the U.S. Department of Energy, Office of
Science, Office of Nuclear Physics, under contract No. DE- SC0012704,
and in part by Laboratory Directed Research and Development (LDRD)
funds from Brookhaven Science Associates.

\end{document}